\documentclass[12pt]{article}
\pdfoutput=1
\usepackage{putex}
\usepackage{feyn}
\usepackage[vcentermath]{youngtab}
\usepackage{subfig}
\usepackage{lscape}

\usepackage{graphicx}
\usepackage{epstopdf}
\usepackage{enumerate}
\usepackage{cite}
\usepackage{tensor}
\usepackage{slashed}
\usepackage{amsmath}
\usepackage{amssymb}
\usepackage{mathrsfs}
\usepackage{lgrind}

\usepackage{bbm}

\usepackage{hyperref}

\numberwithin{equation}{section}

\newcommand {\be} {\begin {equation}}
\newcommand {\ee} {\end {equation}}

\newcommand {\bes} {\begin {equation*}}
\newcommand {\ees} {\end {equation*}}

\newcommand{\eps}{\epsilon}

\def\CH{{\cal H}}

\newcommand{\beq}{\begin{equation}}
\newcommand{\eeq}{\end{equation}}

\def\be{ \begin{equation} }
\def\ee{ \end{equation} }

\def\Tr{{\textrm{Tr}}}
\begin{document}

\preprint{PUPT-2499}

\institution{PU}{Department of Physics, Princeton University, Princeton, NJ 08544}
\institution{PCTS}{Princeton Center for Theoretical Science, Princeton University, Princeton, NJ 08544}

\title{
On $C_{J}$ and $C_{T}$ in Conformal QED}

\authors{Simone Giombi,\worksat{\PU}  Grigory Tarnopolsky\worksat{\PU} and Igor R.~Klebanov\worksat{\PU,\PCTS}  
}

\abstract{QED with a large number $N$ of massless fermionic degrees of freedom has a conformal phase in a range of space-time dimensions. 
We use a large $N$ diagrammatic approach to calculate the leading corrections to $C_T$, the coefficient of the two-point function of the stress-energy tensor, and $C_J$,
the coefficient of the two-point function of the global symmetry current. We present explicit formulae as a function of $d$ and check them versus the expectations in 2 and
$4-\epsilon$ dimensions. Using our results in higher even dimensions we find a concise formula for $C_T$ of the conformal Maxwell theory with higher derivative action
$F_{\mu \nu} (-\nabla^2)^{\frac{d}{2}-2} F^{\mu \nu}$.
In $d=3$, QED has a topological symmetry current, and we calculate the correction to its two-point function coefficient,  
$C^{\textrm{top}}_{J}$. We also show that some RG flows involving QED in $d=3$ obey $C_T^{\rm UV} > C_T^{\rm IR}$ and discuss 
possible implications of this inequality for the symmetry breaking at small values of $N$.
}

\date{}
\maketitle

\tableofcontents

\section{Introduction and Summary}

One of the important observables in Conformal Field Theory (CFT) is $C_T$, the coefficient of the two-point function of the stress-energy tensor  $T_{\mu\nu}$, defined via \cite{Osborn:1993cr}
\begin{align}
\langle T_{\mu\nu}(x_{1}) T_{\lambda \rho}(x_{2})\rangle = C_{T} \frac{I_{\mu\nu,\lambda\rho}(x_{12})}{(x_{12}^{2})^{d}}
\label{TTx}
\ ,\end{align}
where
\begin{align}
& I_{\mu\,\nu,\lambda\rho}(x) \equiv \frac{1}{2}(I_{\mu\lambda}(x)I_{\nu\rho}(x)+I_{\mu\rho}(x)I_{\nu\lambda}(x))-\frac{1}{d}\delta_{\mu\nu}\delta_{\lambda\rho}
\,, \notag\\
& I_{\mu\nu}(x)\equiv \delta_{\mu\nu}-2\frac{x_{\mu}x_{\nu}}{x^{2}}  \,.
\end{align}
If the CFT has a global symmetry generated by conserved currents $J_\mu^a$, then another interesting observable is $C_J$, the coefficient of their two-point functions:
\begin{align}
\label{JJx}
\langle J_{\mu}^{a}(x_{1}) J_{\nu}^{b}(x_{2})\rangle = C_{J} \frac{I_{\mu\nu}(x_{12})}{(x_{12}^{2})^{d-1}} \delta^{ab}\, .  
\end{align}
In CFTs with a large number of degrees of freedom, $N$, these observables typically admit $1/N$ expansions of the form
\begin{align}
&C_{J} = C_{J0}\Big(1+\frac{C_{J1}}{N}+\frac{C_{J2}}{N^{2}}+\mathcal{O}(1/N^{3})\Big)\,,\notag\\
&C_{T} = C_{T0}\Big(1+\frac{C_{T1}}{N}+\frac{C_{T2}}{N^{2}}+\mathcal{O}(1/N^{3})\Big)\,.
\label{CJCT-exp}
\end{align}
The values of $C_{J1}$ and $C_{T1}$ have been calculated in a variety of models. Petkou \cite{Petkou:1995vu} has used large $N$ methods and operator products expansions to calculate
them as a function of $d$ in the scalar $O(N)$ model. Very recently, these results were reproduced using the large $N$ diagrammatic approach in
\cite{Diab:2016spb},
where the same technique was also used to calculate $C_{J1}$ and $C_{T1}$ as a function of $d$ in the conformal Gross-Neveu model. 
An important feature of the diagrammatic approach, which was uncovered in \cite{Diab:2016spb}, is the necessity, in the commonly used 
regularization scheme \cite{Vasiliev:1975mq, Vasiliev:1981yc, Vasiliev:1981dg, Derkachov:1997ch, Ciuchini:1999cv}, of a divergent
multiplicative ``renormalization" $Z_T$ for the stress-energy tensor. 
This factor is required by 
the conformal Ward identities in the regularized theory.  



In this 
paper we extend the methods of \cite{Diab:2016spb} to calculate $C_{J1}(d)$ and $C_{T1}(d)$ in the conformal QED in $d$ dimensions.
This theory, which is reviewed in section \ref{QEDd}, may be thought of as the Maxwell field coupled to $N_f$ massless 4-component Dirac fermions continued from 4 dimensions to a more general dimension $d$. The large $N$ expansion in this model runs in powers of the total 
number of fermionic components, which is $N=4 N_f$. In the physically interesting dimension $d=3$, this 
corresponds to an even number $2N_f$ of two-component Dirac fermions. 

Our main results are
\begin{align}
C_{J1}(d) &=\eta_{m1} \left(\frac{3 d(d-2)}{8 (d-1)}\Theta(d )+\frac{d -2}{d }\right)\,, \label{CJ1answIntr} \\
C_{T1}(d) & = \eta_{m1}\left(\frac{3d (d-2)}{8 (d -1)} \Theta(d)+\frac{d  (d-2)}{(d-1) (d+2)}\Psi(d)-\frac{(d-2) (3 d^2+3 d-8)}{2 (d-1)^2 d (d+2)}\right)\,, \label{CT1answIntr}\\
\Theta(d)&\equiv \psi'(d/2)-\psi'(1)\,,\qquad \Psi(d)\equiv \psi(d-1)+\psi(2-d/2)-\psi(1)-\psi(d/2-1)\,, \notag
\end{align}
where $\psi(x)=\Gamma'(x)/\Gamma(x)$.
Here $\eta_{m1}(d)$ encodes the electron mass anomalous dimension; it is \cite{Gracey:1993sn}\footnote{We define the anomalous dimension of the electron mass operator $O_{m}=\bar{\psi}\psi$ as $\Delta_{O_{m}}=d-1+\eta_{m}$, where $\eta_{m}=\eta_{m1}/N+\mathcal{O}(1/N^{2})$.}
\begin{align}
\eta_{m1}(d) = -\frac{2(d -1) \Gamma (d )}{ \Gamma (\frac{d}{2} )^2 \Gamma (\frac{d}{2}+1) \Gamma (2-\frac{d}{2} )}\,. \label{eta1intr}
\end{align}
In the physically interesting case of $d=3$ we find
\begin{align}
C_{J1}(3) &=\frac{736}{9 \pi ^2}-8\approx 0.285821\, ,\notag \\ 
C_{T1}(3) &=\frac{4192}{45 \pi ^2}-8\approx 1.43863\,. \label{CJCTresD3}
\end{align}

Let us compare our results with the earlier diagrammatic calculations \cite{Huh:2013vga,Huh:2014eea}, which were carried out in $d=3$ using
a regulator different from ours. 
Our result for $C_{J1}(3)$ agrees with that given by Huh and Strack in \cite{Huh:2014eea}.\footnote{ In \cite{Huh:2014eea} only a numerical value
$\tilde{C}_{J}^{(1)}\approx 0.59322699$ was given. We have found the exact expression behind this number: $\tilde{C}_{J}^{(1)}=\frac{136}{3 \pi ^2}-4$,
which leads to $C_{J1}^{\textrm{H\&S}}=\frac{368}{9 \pi ^2}-4$.
The relative factor of 2 between this and our (\ref{CJCTresD3}) is due to the
different conventions: in \cite{Huh:2014eea} $N_f^{\textrm{H\&S}}$ 
is the number of $d=3$ 
Dirac doublets. Therefore, our $N=4 N_f= 2 N_f^{\textrm{H\&S}}$.}
However, our value of  $C_{T1}(3)$ does not agree with that given in \cite{Huh:2014eea}, which after translating to our convention for $N$ is  
$C^{\textrm{H\&S}}_{T1} =  \frac{3808}{45\pi^{2}}-8 \approx 0.574024$.\footnote{ In \cite{Huh:2014eea} only a numerical value
$\tilde{C}_{T}^{(1)}\approx -0.41548168$ was given. We have found the exact expression behind this number: $\tilde{C}_{T}^{(1)}= \frac{1592}{45 \pi^{2}}-4$.}
The source of the disagreement 
is the effect of $Z_T$, which was not included in
\cite{Huh:2014eea}.

A nontrivial check of our results (\ref{CJ1answIntr}) and
(\ref{CT1answIntr})
comes from comparing them with the known exact values in $d=2$
and the $4-\epsilon$ expansions, see sections \ref{CJ1-CT1} and \ref{4minuseps}. Had we not included $Z_T$, there would be no agreement with the
$4-\epsilon$ expansion. 
In higher even $d$, the conformal QED reduces to a free theory of $N$ fermions and 
a conformal higher-derivative Maxwell theory with the action (see e.g. \cite{Giombi:2015haa})
\begin{equation}
F_{\mu \nu} (-\nabla^2)^{\frac{d}{2}-2} F^{\mu \nu}\ .
\label{induced}
\end{equation}
Using the value of $C_{T1}$ in general even dimensions, we extract the $C_T$ of this conformal Maxwell theory
\begin{equation}
C_{T}^{\textrm{conf.\ Maxwell}}|_{{\rm even }\ d} 
=(-1)^{\frac{d}{2}} \frac{d}{ S_d^2} \begin{pmatrix}d\\ \frac{d}{2}-1\end{pmatrix}
\ ,
\end{equation}
where $S_d=\frac{2 \pi^{d/2}}{\Gamma(d/2)}$.

In $d=3$ the QED has a special ``topological" $U(1)$ symmetry current $j^{\rm top}=\frac{1}{2\pi}*F$. In section \ref{toposection} we calculate its 
two-point function to order $1/N^2$, and obtain the associated $C_J^{\rm top}$ coefficient, in the normalization (\ref{JJx}), to be 
\begin{equation}
C_J^{\rm top} = \frac{16}{\pi^{4}N}\left(1+\frac{1}{N}\Big(8-\frac{736}{9 \pi ^2}\Big)+\mathcal{O}(1/N^{2})\right)\,,
\label{CJtop}
\end{equation}
where $N=4N_f$ is twice the number of two-components Dirac fermions. The leading order term is in agreement with \cite{Huh:2013vga, Chester:2016wrc}. 

The QED$_3$ Lagrangian also has an enhanced $SU(2N_f)$ global symmetry, and for small $N_f$
this symmetry may be broken spontaneously to $SU(N_f)\times SU(N_f)\times U(1)$ \cite{Pisarski:1984dj,Appelquist:1988sr}. 
In section \ref{constraint} we present a new estimate for the critical value of $N_f$ above which the symmetry breaking cannot occur 
by using the RG inequality  $C_T^{\textrm{UV}} > C_T^{\textrm{IR}}$.
It implies that the chiral symmetry cannot be broken for $N_{f} > 1+\sqrt 2$. The status of this conclusion is uncertain, since there are known violations of the
inequality in some supersymmetric RG flows \cite{Nishioka:2013gza}. Nevertheless, it is interesting that the critical value of $N_f$ it yields is close to 
other available estimates \cite{Kaveh:2004qa,Fischer:2004nq,Braun:2014wja, Zohar, Giombi:2015haa} and is consistent with the results available from lattice gauge theory
\cite{Strouthos:2008kc,Raviv:2014xna}. 

\section{Large $N$ Expansion for Conformal QED$_d$}
\label{QEDd}

The action for Maxwell theory coupled to $ N_f$ massless charged fermions in flat Euclidean space
\begin{equation}
S = \int d^d x \bigg(\frac{1}{4e^2}F^{\mu\nu}F_{\mu\nu} -\sum_{i=1}^{N_f} \bar\psi_{i} \gamma^{\mu}(\partial_{\mu}+i A_{\mu})\psi^{i} \bigg)\,.
\end{equation}
Here the fermions $\psi^i$ are taken to be four-component complex spinors.  We define the dimensional continuation of the theory by keeping
the number of fermion components fixed. In other words, we take $\gamma^{\mu}$ to be $4\times 4$ matrices satisfying $\{\gamma^{\mu},\gamma^{\nu}\} = 2\delta^{\mu\nu}\, {\bf 1}$, with $\Tr {\bf 1} = 4$. All vector indices are formally continued to $d$ dimensions, i.e. 
$\delta^{\mu\nu}\delta_{\mu\nu}=d$, $\gamma^{\mu}\gamma_{\mu} = d\cdot{\bf 1}$, etc.

One may develop the $1/N$ expansion of the theory
by integrating out the fermions \cite{Appelquist:1981vg,Appelquist:1988sr}. 
This produces an effective action for the gauge field of the form
\begin{equation}
S_{\rm eff} = \int d^d x \frac{1}{4e^2}F^{\mu\nu}F_{\mu\nu}+\int d^d x d^d y\left(\frac{1}{2} A^{\mu}(x)A^{\nu}(y)\langle J_{\mu}(x)J_{\nu}(y)\rangle_0+
\mathcal{O}(A^3)\right)\,,
\label{Seff}
\end{equation}
where
\begin{equation}
J_{\mu} = \bar\psi_{i} \gamma_{\mu}\psi^{i}
\end{equation}
is the conserved $U(1)$ current. Using the bare fermion propagator
\begin{equation}
\delta^i_jG(p)=\langle \psi^{i}(p)\bar\psi_{j}(-p)\rangle 
= \delta^i_j   \frac{i\slashed{p}}{p^2} \,,
\end{equation}
where $\slashed{p} \equiv \gamma^{\mu}p_{\mu}$,
the current two-point function in the free fermion theory is found to be
\begin{equation}
\langle J_{\mu}(p)J_{\nu}(-p)\rangle_0 =
N \frac{ 2\Gamma (2-\frac{d}{2} ) \Gamma (\frac{d}{2} )^2}{ (4\pi) ^{\frac{d}{2}}\Gamma (d )}\left(\delta_{\mu\nu}-\frac{p_{\mu}p_{\nu}}{p^2}\right)(p^2)^{\frac{d}{2}-1}\,.
\label{A2-induced}
\end{equation}
Thus, when $d<4$, one sees that the non-local kinetic term in (\ref{Seff}) is dominant in the low momentum (IR) limit compared to the two-derivative
Maxwell term. Hence, the latter can be dropped at low energies, and one may develop the $1/N$ expansion of the critical theory by
using the induced quadratic term
\begin{equation}
S_{\textrm{crit QED} } =\int \frac{d^dp}{(2\pi)^d}\bigg(  \frac{1}{2}A^{\mu}(p)\langle J_{\mu}(p)J_{\nu}(-p)\rangle_0 A^{\nu}(-p) -\bar\psi_{i} \,i\slashed{p}\,\psi^{i} -i\bar\psi_{i} \gamma^{\mu}A_{\mu}\psi^{i} \bigg)\,.
\label{ScritQED}
\end{equation}
Note that this effective action is gauge invariant as it should, due to conservation of the current.

The induced photon propagator is obtained by inverting the non-local kinetic term in (\ref{ScritQED}). As usual, this requires gauge-fixing. 
Working in a generalized Feynman gauge, the propagator is
\begin{align}
D_{\mu\nu}(p)=  \frac{C_{A}}{N(p^{2})^{\frac{d}{2}-1+\Delta}}\Big(\delta_{\mu\nu}-(1-\xi)\frac{p_{\mu}p_{\nu}}{p^{2}}\Big)\,,
\label{A-prop}
\end{align}
where $\xi$ is an arbitrary gauge parameter ($\xi=0$ corresponds to Landau gauge $\partial_{\mu}A^{\mu}=0$). The normalization constant $C_A$ is 
given by
\begin{align}
C_{A}= \frac{(4\pi)^{\frac{d}{2} }\Gamma (d )}{ 2 \Gamma (\frac{d}{2} )^2\Gamma (2-\frac{d}{2} )} \label{CA}
\end{align}
and in (\ref{A-prop}) we have introduced, as in \cite{Diab:2016spb}, a regulator $\Delta$ to handle divergences
\cite{Vasiliev:1975mq, Vasiliev:1981yc, Vasiliev:1981dg, Derkachov:1997ch, Ciuchini:1999cv}, which should be sent 
to zero at the end of the calculation. This makes the interaction vertex in (\ref{ScritQED}) dimensionful, and one should introduce a renormalization 
scale $\mu$ so that $S_{\rm vertex} = -i\mu^{\Delta} \int \bar\psi_{i} \gamma^{\mu}A_{\mu}\psi^{i}$. 

\begin{figure}[h!]
   \centering
\includegraphics[width=15cm]{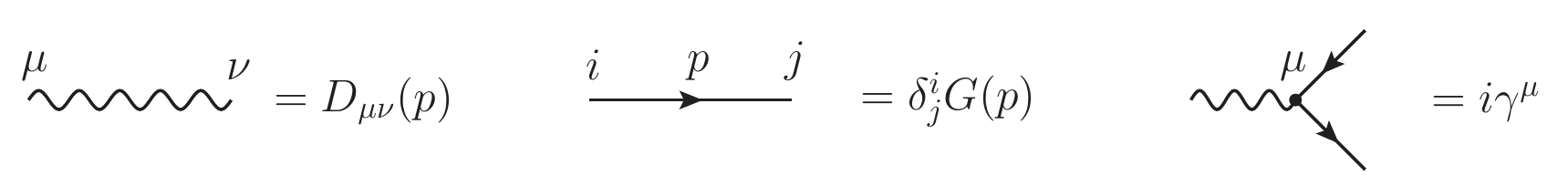}
\caption{Feynman rules for the Large $N$ QED .}
\label{FeynRules}
\end{figure}

The Feynman rules of the model are summarized in figure \ref{FeynRules}. 
In what follows we calculate the two-point function of the $SU(N_{f})$ current and stress-energy tensor, which are given by\footnote{As it was pointed out in \cite{Nielsen:1977sy}, for correlation functions with only gauge invariant operators we can omit the gauge fixing part and ghost part of the stress-energy tensor. This was explicitly checked in QCD in $d=4$ up to three-loops in \cite{Zoller:2012qv}.  } 
\begin{align}
&J_{\nu}^{a} =  -\bar{\psi}_{i}(t^{a})^{i}_{j}\gamma_{\nu}\psi^{j}\,, \notag\\
&T_{\mu\nu} =  -\frac{1}{4}\big(\bar{\psi}_{i}\gamma_{(\mu}D_{\nu)}\psi^{i} - D^{*}_{(\mu}\bar{\psi}_{i} \gamma_{\nu)} \psi^{i}\big)\,,
\label{Tcrit}
\end{align}
where $\gamma_{(\mu}D_{\nu)}\equiv \gamma_{\mu}D_{\nu}+\gamma_{\nu}D_{\mu}$ and $D_{\mu}=\partial_{\mu}+iA_{\mu}$. Note that there is no Maxwell 
term contribution in $T_{\mu\nu}$, as this term was dropped in (\ref{ScritQED}) in the critical limit. 

We will work in flat Euclidean $d$-dimensional metric and introduce a null vector $z^{\mu}$, which satisfies
\begin{align}
z^{2}= z^{\mu}z^{\nu}\delta_{\mu\nu}=0\ .
\end{align}
From (\ref{TTx}), (\ref{JJx}), we see that the two-point functions of the projected operators 
$T \equiv z^{\mu} z^{\nu}T_{\mu\nu}$ and $J \equiv z^{\mu}J_{\mu}$ have the form
\begin{align}
&\langle T (x)T (0)\rangle = \frac{4C_T}{(x^{2})^{d}}\frac{x_z^4}{x^4} \,, \notag\\
&\langle J^a (x) J^b (0)\rangle = \delta^{ab}\frac{-2C_J}{(x^{2})^{d-1}}\frac{x_z^2}{x^2}\,,
\label{twopointfun}
\end{align}
where we have introduced the notation $x_z \equiv z^{\mu}x_{\mu}$. It will be also useful to report the form of these two-point functions in 
momentum space, which may be obtained by Fourier transform and reads
\begin{align}
&\langle T(p)T(-p) \rangle = C_{T} \frac{\pi ^{\frac{d}{2}  } \Gamma (2-\frac{d}{2}  )}{2^{d -2} \Gamma (d +2)}\,\frac{p_{z}^{4}}{(p^{2})^{2-\frac{d}{2} }}\,, \notag\\
&\langle J^{a}(p)J^{b}(-p) \rangle=C_{J} \frac{\pi ^{\frac{d}{2} } \Gamma (2-\frac{d}{2}  ) }{2^{d -3} \Gamma (d )}\,\frac{p_{z}^{2}}{(p^{2})^{2-\frac{d}{2} }}\delta^{ab}\ ,
\label{TTJJp}
\end{align}
where $p_{z}\equiv z^{\mu}p_{\mu}$.

\begin{figure}[h!]
   \centering
\includegraphics[width=17cm]{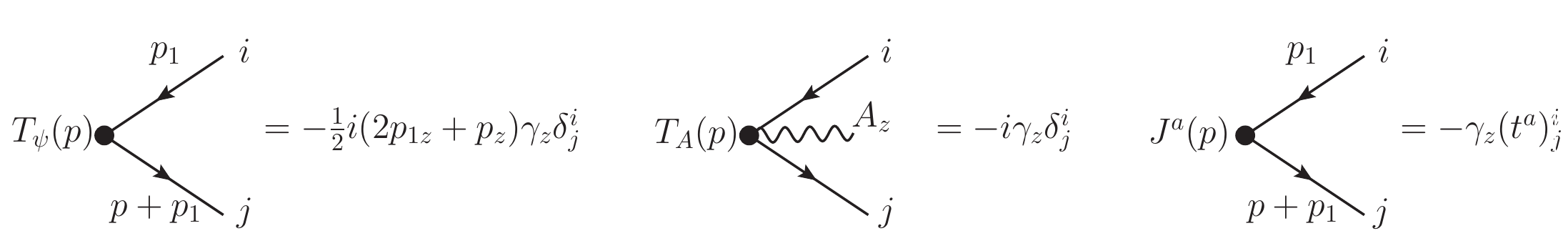}
\caption{Diagramatic representation for $T=T_{\psi}+T_{A}$ and $J^{a}$.}
\label{TJQEDdiag}
\end{figure}

For the stress-tensor of conformal QED, we may write $T=T_{\psi}+T_{A}$, where the two terms are given in 
momentum space by
\begin{align}
T_{\psi}(p)&= -\frac{1}{2}\int \frac{d^{d}p_{1}}{(2\pi)^{d}} \bar{\psi}_{i}(-p_{1})i\gamma_{z}(2p_{1z}+p_{z})\psi^{i}(p+p_{1})\,, \notag\\
T_{A}(p)&=- \int \frac{d^{d}p_{1}}{(2\pi)^{d}} \bar{\psi}_{i}(-p_{1})i\gamma_{z} A_{z}\psi^{i}(p+p_{1})\,, \notag\\ 
J^{a}(p)&=-\int \frac{d^{d}p_{1}}{(2\pi)^{d}} \bar{\psi}_{ i}(-p_{1})(t^{a})^{i}_{j}\gamma_{z}\psi^{j}(p+p_{1})\,.
\end{align}
The diagrammatic representation is shown in figure \ref{TJQEDdiag}.

\section{Calculation of $C_{J1}$ and $C_{T1}$}
\label{CJ1-CT1}

The diagrams contributing to $\langle JJ\rangle$ up to order $1/N$ 
\begin{align}
\langle J^{a}(p)J^{b}(-p)\rangle  = D_{0}+D_{1}+D_{2}+\mathcal{O}(1/N^{2})\,
\end{align}
are shown in figure \ref{CJQED}. Their expressions in momentum space and explicit results are listed in Appendix \ref{apb}. 
\begin{figure}[h!]
   \centering
\includegraphics[width=14cm]{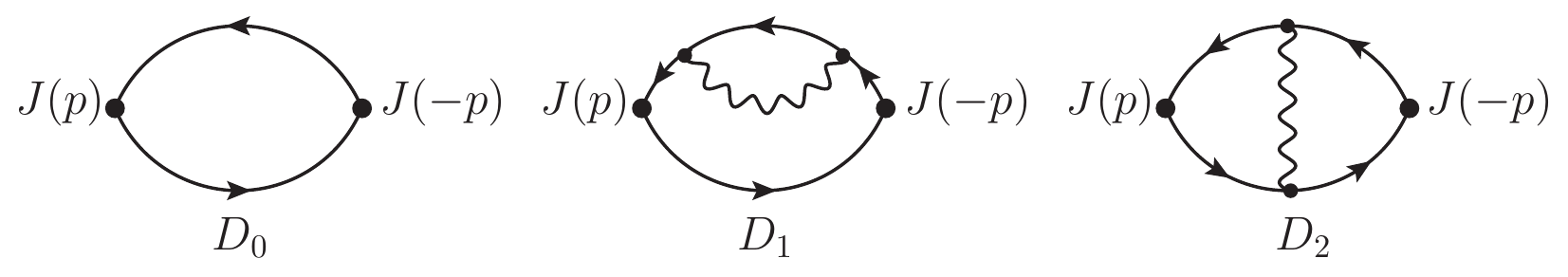}
\caption{Diagrams contributing to $C_{J}$ up to  order $1/N$.}
\label{CJQED}
\end{figure}

\begin{figure}[h!]
   \centering
\includegraphics[width=6cm]{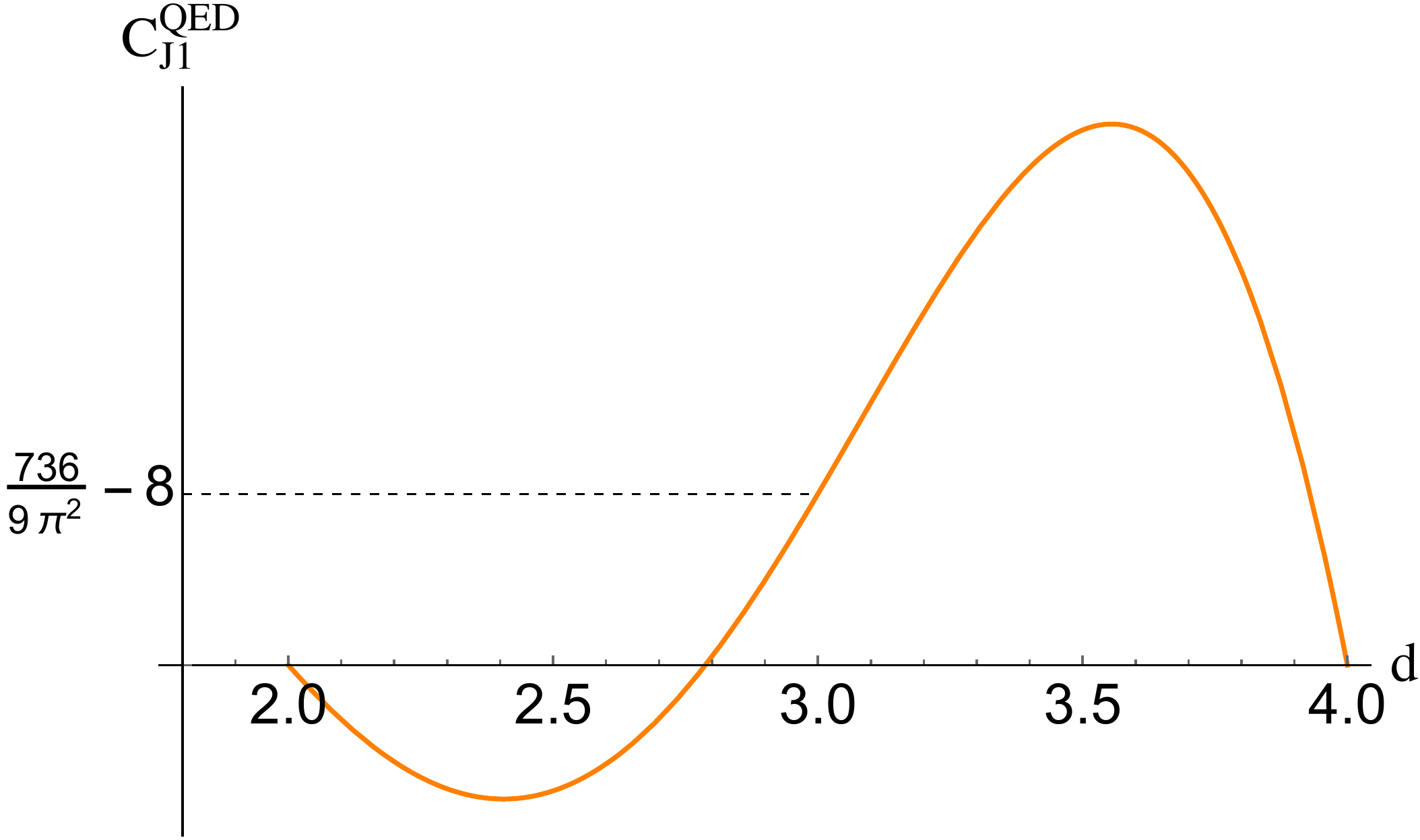}
\caption{Plot of $C_{J1}$.}
\label{CJgraph}
\end{figure}

Putting together the results, we find 
\begin{equation}
\langle J^{a}(p)J^{b}(-p)\rangle = -\tr (t^a t^b)\,
C_{J0}\left(1+\frac{C_{J1}(d)}{N}+\mathcal{O}(1/N^2)\right)\,
 \frac{\pi ^{\frac{d}{2} } \Gamma (2-\frac{d}{2}  ) }{2^{d -3} \Gamma (d )}\frac{p_{z}^{2}}{(p^{2})^{2-\frac{d}{2} }}\,,
\end{equation}
where $C_{J1}(d)$ is given in (\ref{CJ1answIntr}), and
\begin{equation}
 C_{J0} = \Tr {\bf 1} \frac{1}{S_d^2}
\label{CJ-free-ferm}
\end{equation}
is the free fermion contribution. 
A plot of $C_{J1}$ as a function of $d$ is given in figure \ref{CJgraph}.
The value in $d=3$ was given in (\ref{CJCTresD3}) above. One may also extract the following $\epsilon$-expansions 
\begin{align}
C_{J1}|_{d=2+\epsilon}=-\epsilon+\mathcal{O}(\eps^{2}) ,\qquad C_{J1}|_{d=4-\epsilon}= \frac{9 \epsilon }{2}+\left(\frac{9}{2}-9 \zeta (3)\right) \epsilon ^2+\mathcal{O}(\eps^{3})\,. \label{CJ1answ2}
\end{align}
In $d=3$ the leading correction is quite small even for small $N$; for $N=4$, corresponding to $N_f=1$, it makes $C_J$ around $7 \%$ bigger than the free fermion result. 

\begin{figure}[h!]
   \centering
\includegraphics[width=18cm]{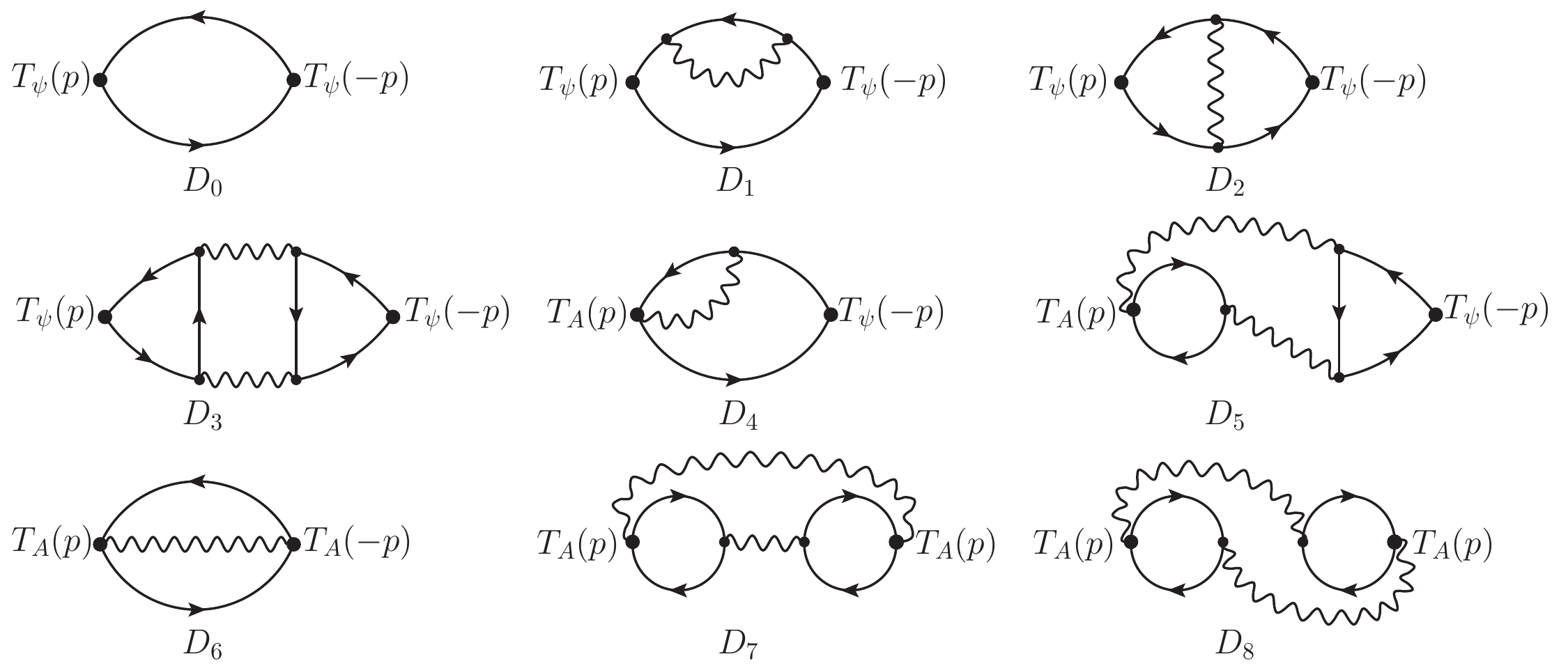}
\caption{Diagrams contributing to $C_{T}$ up to $N^{0}$ order.}
\label{CTQEDdiag}
\end{figure}

Let us now turn to the calculation of $C_T$. Up to order $N^0$, the stress-tensor two-point function receives contribution from 
the diagrams shown in figure \ref{CTQEDdiag}. Note that for some topologies we did not draw explicitly diagrams with the opposite fermion loop direction, but they have to be included. We list the integrands and results for these diagrams in Appendix  B.
We have
\begin{align}
\langle T^{\textrm{ren}}(p) T^{\textrm{ren}}(-p)\rangle = Z_{T}^{2}\langle T(p) T(-p)\rangle  = Z_{T}^{2}\Big(\sum_{n=0}^{8}D_{n}+\mathcal{O}(1/N)\Big)\,,
\end{align}
where we have introduced a ``$Z_{T}$-factor" \cite{Diab:2016spb}, which 
is computed in Appendix A from the Ward identity. It reads $Z_{T}=1+(Z_{T1}/\Delta+Z'_{T1})/N+\mathcal{O}(1/N^{2})$, with 
\begin{align}
Z_{T1} =-\frac{ d (d -2) \eta_{m1}}{2(d +2) (d -1)},\qquad Z'_{T1}=-\frac{ (d -2) \eta_{m1}}{(d +2) (d -1)} \,,  \label{Ztfermap}
\end{align}
where $\eta_{m1}$ is given in (\ref{eta1intr}). Putting together the results for the diagrams given in Appendix \ref{apb}, we obtain 
\begin{equation}
\langle T^{\textrm{ren}}(p) T^{\textrm{ren}}(-p)\rangle  = C_{T0}\left(1+\frac{C_{T1}(d)}{N}+\mathcal{O}(1/N^2)\right)
\frac{\pi ^{\frac{d}{2}  } \Gamma (2-\frac{d}{2}  )}{2^{d -2} \Gamma (d +2)}\frac{p_{z}^{4}}{(p^{2})^{2-\frac{d}{2} }}\,,
\end{equation}
where $C_{T1}(d)$ is given in (\ref{CT1answIntr}), and the free fermion contribution is
\begin{equation}
C_{T0} = N \frac{d}{2S_d^2}\,.
\label{freeferm}
\end{equation}
As a check of our calculation, we note that the final result does not depend on the gauge parameter $\xi$. 

\begin{figure}[h!]
   \centering
\includegraphics[width=8cm]{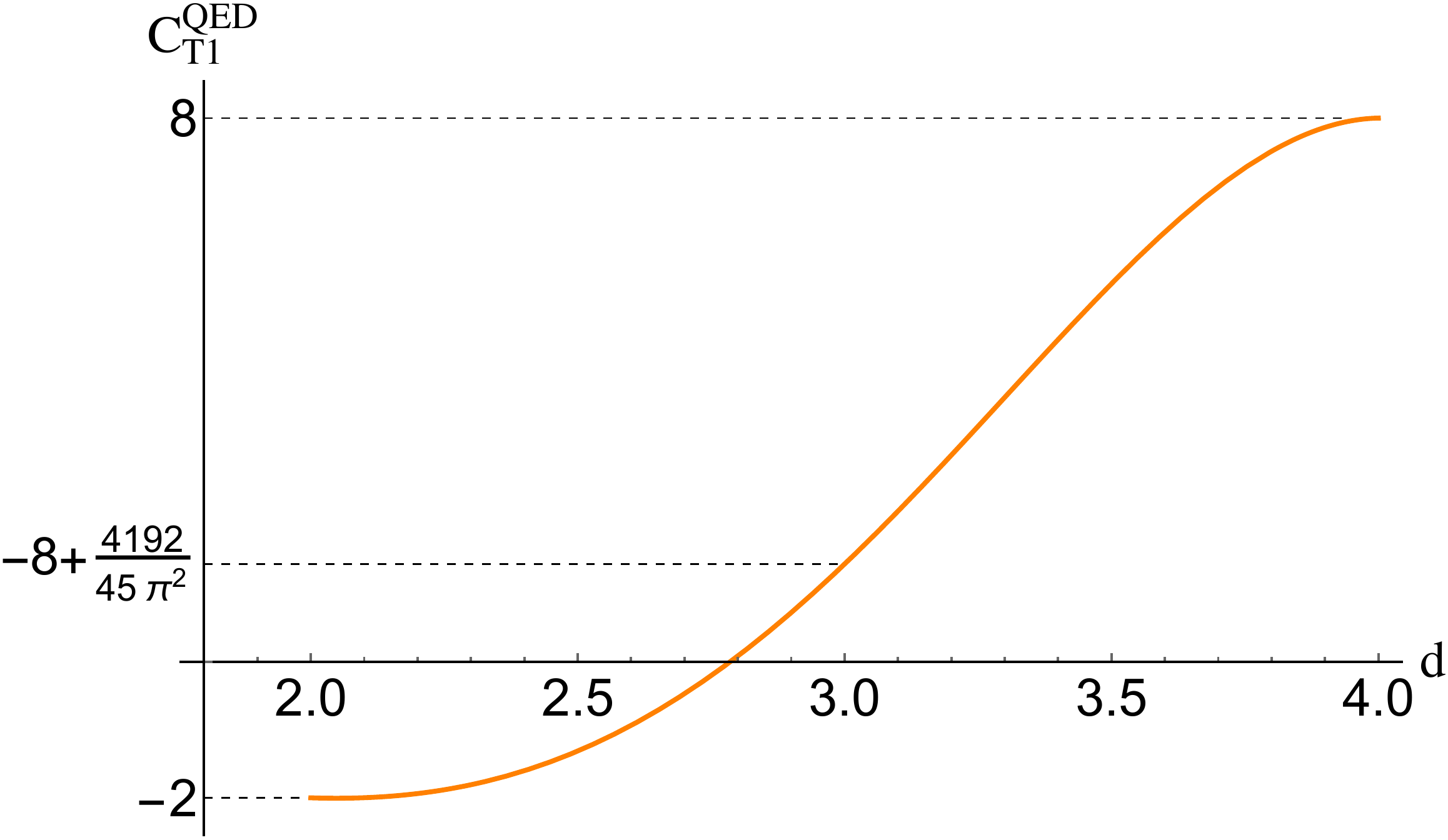}
\caption{Plot of $C_{T1}$.}
\label{CTgraph}
\end{figure}

A plot of $C_{T1}(d)$ in $2<d<4$ is given in figure \ref{CTgraph}.
We see that $C_{T1}$ is negative for $2<d< 2.79$. This means that the inequality $C_T^{\textrm{UV}} > C_T^{\textrm{IR}}$ is violated for the flow from conformal QED$_d$ (which may be thought of as the
UV fixed point of the Thirring model) to the free fermion theory for $2<d< 2.79$. However, it holds for $2.79 < d < 4$, including in particular $d=3$.

Near some even dimensions we find
\begin{align}
C_{T1}|_{d=2+\epsilon}=-2-\frac{\epsilon}{4},\qquad C_{T1}|_{d=4-\epsilon}=8- \frac{\epsilon }{6} ,\qquad C_{T1}|_{d=6-\epsilon}=-30+\frac{61\epsilon }{6}\, .\label{CT1answ2}
\end{align}
Note that in $d=2$ we get 
\begin{equation}
C_T|_{d=2} = \frac{N}{S_2^2}\left(1-\frac{2}{N}\right)\,.
\label{CT2d}
\end{equation}
This result is precisely as expected, since the conformal QED$_2$
corresponds to the multiflavor Schwinger model with $2N_f$ Dirac fermions, which is described by a CFT with central charge $c=2N_f-1$ 
\cite{Gepner:1984au, Affleck:1985wa}. Normalizing 
(\ref{CT2d}) by the free scalar contribution $C_{T}^{\rm sc} = d/((d-1)S_d^2)$, and recalling $N=4N_f$, we obtain precisely this central charge. In 
section \ref{4minuseps} we will see that $ C_{T1}|_{d=4-\epsilon}$ also agrees with the $4-\eps$ expansion.  


Near
even dimensions the QED$_d$ theory is expected to be described by the free fermions weakly coupled to a $U(1)$ gauge theory with the local kinetic term (\ref{induced}).
For example, in $d=6$ this higher-derivative theory was explored in
\cite{Ivanov:2005qf,Ivanov:2005kz,Smilga:2005pr,Smilga:2006ax,Beccaria:2015uta,Giombi:2015haa,Gracey:2015xmw}. We may use  (\ref{CT1answIntr}) to extract the $C_T$ 
coefficient for the conformal Maxwell theory (\ref{induced}).
From (\ref{CT1answIntr}) it follows that
\begin{equation}
C_{T1}^{\textrm{QED}}|_{{\rm even }\ d} = \frac{2(-1)^{\frac{d}{2}}d!}{(\frac{d}{2}-1)!(\frac{d}{2}+1)!}
= 2(-1)^{\frac{d}{2}}\begin{pmatrix}d\\ \frac{d}{2}-1\end{pmatrix}
\ .\label{CTqed-2n}
\end{equation}
Recalling that the contribution of the free massless fermions is given by (\ref{freeferm}), 
we find that the $C_T$ of the conformal Maxwell theory is
\begin{equation}
C_{T}^{\textrm{conf.\ Maxwell}}|_{{\rm even }\ d} =\frac{d}{2 S_d^2}C_{T1}^{\textrm{QED}}|_{{\rm even }\ d}
=(-1)^{\frac{d}{2}} \frac{d}{ S_d^2} \begin{pmatrix}d\\ \frac{d}{2}-1\end{pmatrix}
\ .\label{CTconfMax-2n}
\end{equation}
In $d=4,6,8, 10, \ldots$ this formula gives $16, -90, 448, -2100, \ldots$ times $1/S_d^2$.
In $d=4$ this agrees with the standard answer for the Maxwell theory.
In $d=6, 8, \ldots$, eq.~(\ref{CTconfMax-2n}) gives new results for the values of $C_T$ in the free conformal theory with the higher-derivative action (\ref{induced}). 

\section{$C^{\textrm{top}}_{J}$ for the Topological Current in $d=3$}
\label{toposection}

\begin{figure}[h!] 
   \centering
\includegraphics[width=15cm]{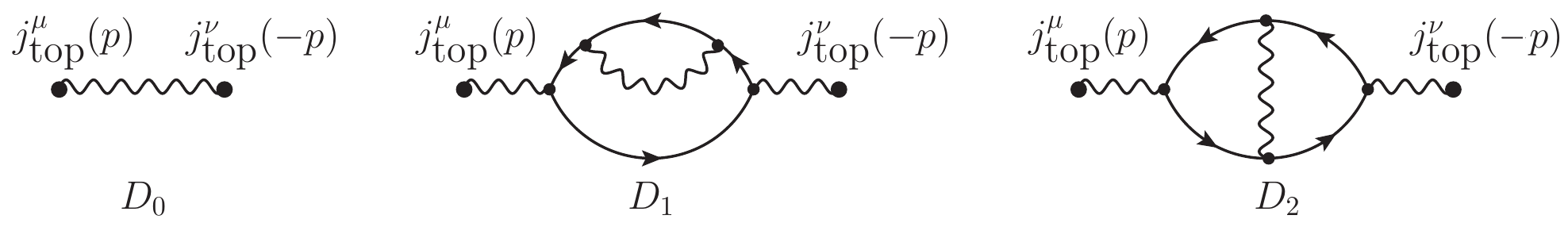}
\caption{Diagrams contributing to $C^{\textrm{top}}_{J}$ up to $1/N^{2}$ order.}
\label{CJtopQED}
\end{figure}

In $d=3$, it is interesting to compute $C^{\textrm{top}}_{J}$ for the ``topological" $U(1)$ current
\begin{equation}
j_{\textrm{top}}^{\mu}= \frac{i}{4\pi}\epsilon^{\mu\nu\lambda}F_{\nu\lambda}\, , \label{jtoddef}
\end{equation}
where the factor of $i$ arises because we are working in Euclidean signature, and the normalization is such that the associated charges are integers. 
The diagrams contributing to the current two-point function up to order $1/N^{2}$ are shown in figure \ref{CJtopQED}.  
The diagrams $D_{1}$ and $D_{2}$ have the same structure as the corresponding ones in fig. \ref{CJgraph} for the $SU(N_{f})$ current,\footnote{In fact these diagrams can be extracted from the evaluation of the polarization operator, which was computed in case of QCD  in \cite{Ciuchini:1999wy}.} with the difference that at the external points 
we now have the gauge $U(1)$ current, to which we attach the two induced photon propagators.  
Thus, using the results from Appendix \ref{apb}, we find  
\begin{align}
\langle j_{\textrm{top}}^{\mu} (p)j_{\textrm{top}}^{\nu} (-p)\rangle &=-\frac{1}{(4\pi)^{2}} \epsilon^{\mu\rho \sigma}e^{\nu\tau \lambda} \langle (p_{\rho}A_{\sigma}(p)-p_{\sigma}A_{\rho}(p))(p_{\tau}A_{\lambda}(-p)-p_{\lambda}A_{\tau}(-p)) \rangle \notag\\
&= -\frac{|p|}{4\pi^{2}} \frac{C_{A}}{N} \Big(1-\frac{C_{J1}(3)}{N}+\mathcal{O}(1/N^{2})\Big)\Big(\delta_{\mu\nu}-\frac{p_{\mu}p_{\nu}}{p^{2}}\Big)\,,
\end{align}
where  $C_{A}$ and $C_{J1}(d)$ are given in (\ref{CA}) and (\ref{CJ1answIntr}), which yield $C_{A}|_{d=3}=32$ and the value of $C_{J1}(3)$ given in (\ref{CJ1answ2}). 
Therefore, we finally get 
\begin{align}
\langle j_{\textrm{top}}^{\mu} (p)j_{\textrm{top}}^{\nu} (-p)\rangle &=-\frac{8 |p|}{\pi^{2}N}\left(1+\frac{1}{N}\Big(8-\frac{736}{9 \pi ^2}\Big)+\mathcal{O}(1/N^{2})\right)\Big(\delta_{\mu\nu}-\frac{p_{\mu}p_{\nu}}{p^{2}}\Big)\,.
\end{align}
Comparing with the momentum space normalization in (\ref{TTJJp}), we find the result given in eq.~(\ref{CJtop}). We note that this is related to $C_J$ in (\ref{CJCT-exp})-(\ref{CJCTresD3}) by an inversion, $C_A \sim 1/C_J$. This essentially follows from the fact that in the large $N$ critical QED, $A_{\mu}$ and $J_{\mu}$ are related by a Legendre transformation \cite{Witten:2003ya,Leigh:2003ez}, see eq.~(\ref{ScritQED}).

The conformal bootstrap constraints on the values of $C_J$, $C_T$ and $C^{\textrm{top}}_{J}$ in QED$_3$ for $N_f=1,2,3$ were recently discussed in
\cite{Chester:2016wrc}. 
In Table \ref{tableCTCJres} we summarize our results for these coefficients in $d=3$ and for 
different values of $N_f$ (the number of 4-component fermions). These results appear to fall within the regions allowed by the bootstrap for $N_f=1,2,3$. 
 \begin{table}[h]
\centering
\begin{tabular}{ccccccccc}
\hline
\multicolumn{1}{|c|}{$N_{f}$}         
& \multicolumn{1}{c|}{1} & \multicolumn{1}{c|}{2} & \multicolumn{1}{c|}{3} & \multicolumn{1}{c|}{4}& \multicolumn{1}{c|}{5} & \multicolumn{1}{c|}{10}& \multicolumn{1}{c|}{20} \\ \hline
\multicolumn{1}{|c|}{$C_{T}/C_{T0}$ }     
& \multicolumn{1}{c|}{1.3597} & \multicolumn{1}{c|}{1.1798} & \multicolumn{1}{c|}{1.1199} & \multicolumn{1}{c|}{1.0899} & \multicolumn{1}{c|}{1.0719}& \multicolumn{1}{c|}{1.0360}  & \multicolumn{1}{c|}{1.0180}\\ \hline
\multicolumn{1}{|c|}{$C_{J}/C_{J0}$ }      
& \multicolumn{1}{c|}{1.0715} & \multicolumn{1}{c|}{1.0357} & \multicolumn{1}{c|}{1.0238} & \multicolumn{1}{c|}{1.0179} & \multicolumn{1}{c|}{1.0143}& \multicolumn{1}{c|}{1.0072}  & \multicolumn{1}{c|}{1.0036}\\ \hline
\multicolumn{1}{|c|}{$8\pi^2\,C_{J}^{\textrm{top}}$ }      
& \multicolumn{1}{c|}{3.0106} & \multicolumn{1}{c|}{1.5632} & \multicolumn{1}{c|}{1.0550} & \multicolumn{1}{c|}{0.7961} & \multicolumn{1}{c|}{0.63919}& \multicolumn{1}{c|}{0.3219}  & \multicolumn{1}{c|}{0.1615}\\ \hline
\end{tabular}
\caption{Results for $C_T$, $C_J$ and $C_J^{\rm top}$ in $d=3$ for different values of $N_f$, the number of 4-component fermions (half the number of 
2-component Dirac spinors). $C_T$ and $C_J$ are normalized by the free field values in (\ref{freeferm}) and (\ref{CJ-free-ferm}). To facilitate the comparison 
with \cite{Chester:2016wrc}, $C_J^{\rm top}$ is normalized by the free fermion contribution (\ref{CJ-free-ferm}) for 2-component spinors ($\Tr {\bf 1}=2$), 
which is $\Tr {\bf 1}/S_3^2 = 1/(8\pi^2)$.}
\label{tableCTCJres}
\end{table}

\section{$4-\epsilon$ Expansion of $C_{J}$ and $C_T$ }
\label{4minuseps}

To find $C_{J}$ in the $4-\eps$ expansion to the leading non-trivial order, we have to compute diagrams with the same topology as those in the large $N$ approach, 
figure \ref{CJQED}, but now the photon propagator is the standard one obtained from the Maxwell term. It reads
\begin{align}
D_{\mu\nu}(p)= \frac{1}{p^{2}}\left(\delta_{\mu\nu}-(1-\xi)\frac{p_{\mu}p_{\nu}}{p^2}\right)\,, \label{qedlandprop}
\end{align}
where we have introduced an arbitrary gauge parameter ($\xi=1$ is the usual Feynman gauge, and $\xi=0$ Landau gauge). 

The renormalization of the electric charge is well-known, and 
in minimal subtraction scheme it reads \cite{Moshe:2003xn}:
\begin{equation}
e_{0} = \mu^{\frac{\epsilon}{2}}\bigg(e+\frac{4  N_{f}}{3  \epsilon }\frac{e^{3}}{(4 \pi )^2}
+\Big(\frac{8  N_{f}^{2}}{3 \epsilon^{2}}+\frac{2 N_{f}}{\epsilon}\Big)\frac{e^5}{(4 \pi )^4}+\ldots \bigg)\,,
\label{e0Toe}
\end{equation}
where $e$ is the renormalized coupling, and the corresponding beta function is
\begin{equation}
\beta=-\frac{\epsilon }{2}e+\frac{4N_{f}}{3}\frac{e^3}{ (4 \pi )^2}+\frac{4 N_{f} e^5 }{(4 \pi )^4}-\frac{2 N_{f} (22 N_{f}+9)}{9 }\frac{e^7}{(4 \pi )^6}+\dots\,.
\label{beta-e}
\end{equation}
Then, one finds an IR stable perturbative fixed point at
\begin{equation}
e_{*}=\pi  \sqrt{\frac{6 \epsilon }{N_{f}}} \bigg(1-\frac{9 }{16 N_{f}}\epsilon +\frac{3 (44 N_{f}+207) }{512 N_{f}^2}\epsilon ^2+\mathcal{O}(\epsilon^3)\bigg)\,.
\label{estar}
\end{equation}
Computing the diagrams in figure \ref{CJQED} with the photon propagator (\ref{qedlandprop}), taking a Fourier transform to coordinate space, and 
setting $e=e_*$ at the end, we obtain in $d=4-\epsilon$
\begin{align}
C_{J}/C_{J}^{\textrm{free}}= 1+\frac{9\epsilon}{8 N_{f}}+\mathcal{O}(1/N_f^2)\,.
\end{align}
which precisely agrees with (\ref{CJ1answ2}) (recall that in this case we have $N= N_{f}\Tr {\bf 1}=4N_{f}$).


To calculate the $4-\epsilon$ expansion of $C_T$ to order $\epsilon$, 
we will use as a shortcut the fact that in $d=4$ the $C_T$ coefficient may be obtained as (see e.g. \cite{Cappelli:1990yc, Petkou:1994ad})
\begin{equation}
C_T= \frac{640}{\pi^2} \beta_a\ ,
\end{equation}
where $\beta_a$ is the beta function for the Weyl-squared term, which is known to be \cite{Hathrell:1981gz, Jack:1990eb}  
\begin{equation}
\beta_a = \frac{N_f+2}{20(4\pi)^2}+\frac{7N_f}{36}\frac{e^2}{(4\pi)^4}+\ldots\,.
\end{equation}
The first term corresponds to the contributions of the free fermions and of the Maxwell field, while the second one encodes the leading interaction corrections.
The second term, when evaluated at the  IR fixed point (\ref{estar}) in $d=4-\epsilon$, gives $\frac{7\epsilon}{6 (16\pi)^2}$.
However, this is not the only contribution of order $\epsilon$ because the free field contributions need to be evaluated in $4-\epsilon$ dimensions. 
The contribution of free massless fermions is given in (\ref{freeferm}).
The contribution of the Maxwell field is more subtle, since this theory is scale invariant but not conformal away from four dimensions \cite{ElShowk:2011gz}. 
However, defining the projected stress-tensor $T_{\textrm{Maxwell}}= z^{\mu} z^{\nu}F_{\mu\alpha} F^\alpha_{\ \nu}$ (this selects 
the traceless part of $T_{\mu\nu}$), and using the field strength 
two-point function \cite{ElShowk:2011gz}
\begin{equation}
\langle F_{\mu\nu}(x) F_{\rho\sigma}(0)\rangle = \frac{(2d-4)\Gamma\left(\frac{d}{2}-1\right)}{4\pi^{\frac{d}{2}} (x^2)^{d/2}} \left[
\left(\delta_{\mu\rho}-\frac{d}{2}x_{\mu}x_{\rho}/x^2\right)\left(\delta_{\nu\sigma}-\frac{d}{2}x_{\nu}x_{\sigma}/x^2\right)
-\mu \leftrightarrow \nu\right]
\end{equation}
we find that $\langle T_{\textrm{Maxwell}}(x)T_{\textrm{Maxwell}}(0)\rangle$ takes the form (\ref{twopointfun}), just as in a conformal field theory, 
with the normalization given by
 \begin{equation}
C_T^{\rm Maxwell}= \frac{d^2 (d-2)}{2 S_d^2}\,.
\label{CTMaxwell}
\end{equation}
This serves as the natural definition of $C_T$ for the Maxwell theory (in $d=4$, it agrees with the well-known result \cite{Osborn:1993cr}). 
Putting these results together we find
\begin{equation}
C_T^{\rm QED}= C_T^{\rm free~ferm} \left (1+ \frac {d(d-2) +35\eps/6}{N} + \ldots \right )= C_T^{\rm free~ferm} \left (1+ \frac {8 -\eps/6}{N} + \ldots \right )\ ,
\end{equation}
which exactly agrees with (\ref{CT1answ2}). This gives a highly non-trivial test of the dimension dependence of $C_{T1}$.


\section{A New Estimate for Symmetry Breaking in QED$_3$}
\label{constraint}

In $d=3$, the QED Lagrangian has $SU(2N_f)$ global symmetry. For $N_f< N_{f, {\rm crit}}$ it may be broken via the generation of vacuum expectation value of the operator 
$\sum_{j=1}^{N_f} \bar \psi_j \psi^j$ (this is written using the 4-component spinors $\psi^i$ and gamma-matrices) \cite{Pisarski:1984dj,Appelquist:1988sr}. This
operator preserves the 3-d time reversal symmetry,
but it breaks the global symmetry to $SU(N_f)\times SU(N_f)\times U(1)$. 

In an earlier paper \cite{Giombi:2015haa}, using the $F$-theorem inequality 
$F^{\textrm{UV}} > F^{\textrm{IR}}$ \cite{Jafferis:2011zi,Myers:2010xs,Casini:2011kv,Klebanov:2011gs,Casini:2012ei} we showed 
that theories with $N_f=5$ and higher must be in the conformal phase. The $F$-theorem method is inconclusive, however, for theories with $N_f\leq 4$. 
There is lattice evidence that theories with  $N_f=1,2$ are not conformal
\cite{Strouthos:2008kc,Raviv:2014xna},\footnote{See, however, the recent lattice work \cite{Karthik:2015sgq} suggesting that they are conformal.} but little is known about theories with $N_f=3,4$.

Let us now consider a different RG inequality: 
\begin{equation}
C_T^{\textrm{UV}} > C_T^{\textrm{IR}}\ ,
\label{ctheorem}
\end{equation}
which is sometimes called ``the $C_T$ theorem".
 While there is a known $d=3$ counter-example to this inequality \cite{Nishioka:2013gza}, which involves theories with
${\cal N}=2$ supersymmetry, many known RG flows appear to obey (\ref{ctheorem}). For example, it is obeyed for flows involving the
scalar $O(N)$ \cite{Petkou:1995vu,Diab:2016spb} and the Gross-Neveu model \cite{Diab:2016spb}. If we think of the conformal QED$_3$ theory as the UV fixed point of the Thirring model,
then the inequality (\ref{ctheorem}) is obeyed by the flow to the free fermion theory because $C_{T1}(3)>0$.
We may also test this inequality for the flow from the QED theory in the extreme UV, which consists of the free decoupled Maxwell field and $N_f$ 4-component fermions, to the
conformal QED$_3$. For the former we find using (\ref{CTMaxwell}) and (\ref{freeferm})
\begin{equation}
C_T^{\textrm{UV}}= \frac{12 N_f +9}{32 \pi^2}.
\label{CTUV}
\end{equation}
For the interacting conformal phase, using our result (\ref{CJCTresD3}), we have
\begin{equation}
C_T^{\textrm{IR}}=\frac{6N_f}{16\pi^2}\left(1+\frac{\frac{4192}{45 \pi ^2}-8}{4N_f}+\mathcal{O}(1/N_f^2)\right)\,.
\end{equation}
We see that at large $N_f$ (\ref{ctheorem}) is obeyed to order $N_f^0$ because 
$9>3\left(\frac{4192}{45 \pi ^2}-8\right) \approx 4.32$. 

Let us now try applying (\ref{ctheorem}) to the $d=3$ flow from QED in the extreme UV 
to the broken symmetry phase. For the former we have (\ref{CTUV}). 
The latter is a free conformal field theory of $2N_f^2+1$ scalar fields; therefore, it has 
\begin{equation}
C_T^{\textrm{IR}}= \frac{3(2N_f^2+1)}{32 \pi^2} .
\end{equation}
We find that the two expressions are equal for $N_f=N_{f, {\rm crit}}=1+ \sqrt{2}\approx 2.414$. This suggests that theories with $N_f=3$ and higher are in the conformal phase.
The inequality (\ref{ctheorem}), however, does not require the $N_f=1,2$ theories to be conformal, and indeed there is lattice evidence that they are not 
\cite{Strouthos:2008kc,Raviv:2014xna}.\footnote{
A more stringent value $N_{f, {\rm crit}}= 3/2$ follows from the RG inequality based on the coefficient of the 
thermal free energy \cite{Appelquist:1999hr}.  
This appears to be in contradiction with
the lattice gauge theory work \cite{Raviv:2014xna} claiming that the $N_f=2$ theory is not conformal. However, both $N_{f, {\rm crit}}= 3/2$
and $N_{f, {\rm crit}}=1+ \sqrt{2}\approx 2.414$
are consistent with the recent paper \cite{Karthik:2015sgq} claiming that the symmetry breaking does not take place even for $N_f=1$.} 

\section{$C_{T}$ for Large $N_{f}$ QCD$_{d}$}

To the leading nontrivial order, the large $N_{f}$ computations for QCD look similar to those in the QED case. The results for large $N_{f}$ QCD at the critical point 
can be deduced from the lagrangian \cite{Hasenfratz:1992jv, Gracey:1993ua, Gracey:2015xmw, Kazakov:2007su, Ali:2001hd, Dudal:2004ch, Ciuchini:1999wy, Ciuchini:1999cv, Bennett:1999he}
\begin{align}
\mathcal{L}_{\textrm{crit~QCD}} =-\bar{\psi}_{i}\gamma^{\mu}(\partial_{\mu}+i A_{\mu}^{a}t^{a})\psi^{i}+\frac{N_{f}}{2\xi} (\Box^{(d-4)/2}\partial A)^{2}+\partial_{\mu}\bar{c}^{a}\partial^{\mu}c^{a}+f^{abc}\partial^{\mu}\bar{c}^{a}A_{\mu}^{b}c^{c}\,,
\label{SQCD}
\end{align} 
where $\psi^{i}$ with $i=1,..,N_{f}$ are the quark fields belonging to the fundamental representation of the colour group $G$, $A_{\mu}^{a}$ is the gluon field and $c^{a}$ and $\bar{c}^{a}$ are the ghost fields in the adjoint representation of the colour group. 
We will use the following notation for the Casimirs of the Lie group generators $t^{a}$ ($[t^{a},t^{b}]=if^{abc}t^{c}$): 
\begin{align}
\tr(t^{a}t^{b}) = C(r)\delta^{ab}, \quad t^{a}t^{a}= C_{2}(r) \cdot I, \quad f^{acd}f^{bcd} = C_{2}(G) \cdot I
\end{align}
and also $\tr (I)=d(r)$ and $\delta^{ab}\delta^{ab}=d(G)$. 
The stress-energy tensor is (\ref{Tcrit})  
with $D_{\mu}=\partial_{\mu}+iA_{\mu}^{a}t^{a}$, and as we mentioned above  we can omit the gauge fixing and  ghost parts of $T_{\mu\nu}$ when computing correlation functions 
of gauge invariant operators. The diagrams contributing to $C_T$ to order $1/N_{f}$ are the same as in the QED case (see figure \ref{CTQEDdiag}).
It is not hard to show that the relations between QED and QCD diagrams are
\begin{align}
D_{0}^{\textrm{QCD}} = d(r) D_{0}^{\textrm{QED}}, \quad D_{n}^{\textrm{QCD}} = d(G) D_{n}^{\textrm{QED}}, \quad n =1,..,8\,,
\end{align}
where for some diagrams we used the identity $d(r)C_{2}(r)=d(G)C(r)$. Therefore, we find 
\begin{align}
C_{T}^{\textrm{QCD}} = d(r) C_{T0} \bigg(1+\frac{1}{N} \frac{d(G)}{d(r)}C_{T1} +\mathcal{O}(1/N^{2})\bigg)\, ,
\label{CTgengauge}
\end{align}
where $C_{T0}$ and $C_{T1}$ are the results for QED given in (\ref{freeferm}) and (\ref{CT1answIntr}).
For  $SU(N_{c})$ gauge group we have $d(r)=N_{c}$ and $d(G)=N_{c}^{2}-1$, thus 
\begin{align}
C_{T}^{\textrm{QCD}} = N_{c}C_{T0} \bigg(1+\frac{1}{N} \frac{N_{c}^{2}-1}{N_{c}}C_{T1} +\mathcal{O}(1/N^{2})\bigg)\,. \label{CTQCDans1}
\end{align}

Let us check that this agrees with the known exact result for central charge in $d=2$ gauge theory with massless flavors.  
The conformal limit of $SU(N_{c})$ gauge theory has central charge \cite{Witten:1983ar,Gepner:1984au,Affleck:1985wa,Bhanot:1993xp}
\begin{align}
c= c_{\textrm{free}} - \frac{(N_{c}^{2}-1)k}{k+N_{c}}\,.
\end{align}
The subtraction of the second term is due to the gauging of the $SU(N_{c})$ Kac-Moody algebra with level $k$.
Since there are $2N_f$ 2-d Dirac flavors in the fundamental representation of $SU(N_{c})$, we have $k=2 N_f$. 
This theory may be described by a $SU(2N_f)_{N_c}\times U(1)$ WZW model \cite{Gepner:1984au,Affleck:1985wa}. Its central charge is
\begin{align}
c= 2 N_f \frac{2 N_f N_c+1}{2N_f + N_c}= N_{f} N_{c}- \frac{2 (N_{c}^2-1) N_f}{2 N_f+N_{c}}= 2 N_f N_{c} \bigg(1- \frac{1}{ 2 N_{f}}\frac {N_{c}^2-1} { N_{c}} + \dots\bigg)\,,
\end{align}
which is in agreement with (\ref{CTQCDans1}) evaluated in $d=2$. For a general gauge group $G$ we have
\begin{align}
c= 2 N_{f}  d(r)  - \frac{2 d(G) N_f }{2 N_f + d(r) }= 2 N_f d(r) \bigg(1- \frac{1}{ 2 N_{f}}\frac {d(G)} { d(r)} + \dots\bigg)\,, 
\end{align}
which agrees with (\ref{CTgengauge}). Analogously, one can easily see that we have the same relation between  $C_{J}$ in QCD and QED:
\begin{align}
C_{J}^{\textrm{QCD}} = d(r)C_{J0} \bigg(1+\frac{1}{N} \frac{d(G)}{d(r)}C_{J1} +\mathcal{O}(1/N^{2})\bigg)\, ,
\label{CJgengauge}
\end{align}
where $C_{J0}$ and $C_{J1}$ are the results for QED given in (\ref{CJ-free-ferm}) and (\ref{CJ1answIntr}).

\bigskip
{\bf Note Added:} After the first version of this paper appeared, the value of $C_T$
for the $d=6$ conformal Maxwell theory was calculated directly in \cite{Osborn:2016bev}. The result is in agreement with our (\ref{CTconfMax-2n}), providing a check of our methods.

\section*{Acknowledgments}

We thank L. Fei and S. Pufu for useful discussions.
The work of SG was supported in part by the US NSF under Grant No.~PHY-1318681.
The work of IRK and GT was supported in part by the US NSF under Grant No.~PHY-1314198.

\appendix

\section{Calculation of $Z_{T}$ } 
\label{apa}

\begin{figure}[h!]
   \centering
\includegraphics[width=16cm]{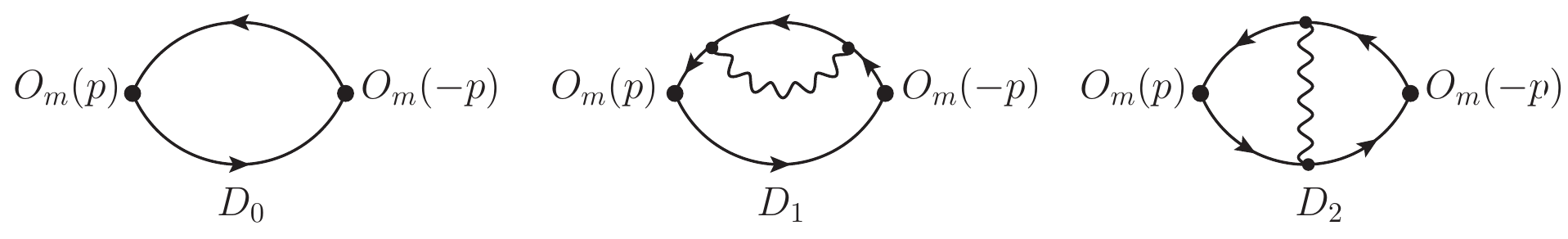}
\caption{Diagrams contributing to $\langle O_{m}(p)O_{m}(-p)\rangle$ up to order $1/N$.}
\label{OmOmDiag}
\end{figure}

In this appendix we present the computation of the $Z_{T}$ factor for the stress-energy tensor in the theory of Critical QED. As we show below, a non-trivial $Z_T$ is required for the Ward identity to hold. We define the ``renormalized'' stress-energy tensor $T^{\textrm{ren}}_{\mu\nu}$ by
\begin{align}
T^{\textrm{ren}}_{\mu\nu}(x) =Z_{T}T_{\mu\nu}(x)\,,
\end{align}
where  $Z_{T}= 1+(Z_{T1}/\Delta+Z'_{T1})/N+\mathcal{O}(1/N^{2})$, and $T_{\mu\nu}$ is the ``bare" stress-tensor. 
To find $Z_{T}$ we will use the three-point function $\langle T^{\textrm{ren}}_{\mu\nu}(x_{1})O^{\textrm{ren}}_{m}(x_{2})O^{\textrm{ren}}_{m}(x_{3})\rangle$,
where $O^{\textrm{ren}}_{m} =Z_{O_{m}} O_{m}$ is the electron mass operator,  $Z_{O_{m}}$ is its renormalization constant and the bare operator is
\begin{align}
O_{m} = \bar{\psi}\psi\,.
 \end{align}
This three point function is gauge invariant.  So using conformal invariance and conservation of the stress-tensor, 
one has the general expression for the three-point  function 
\begin{align}
\langle T^{\textrm{ren}}_{\mu\nu}(x_{1})O^{\textrm{ren}}_{m}(x_{2})O^{\textrm{ren}}_{m}(x_{3}) \rangle =\frac{-C_{TO_{m}O_{m}}}{(x_{12}^{2}x_{13}^{2})^{\frac{d}{2}-1} (x_{23}^{2})^{\Delta_{O_{m}}-\frac{d}{2}+1}} \Big((X_{23})_{\mu}(X_{23})_{\nu}-\frac{1}{d}\delta_{\mu\nu}(X_{23})^{2}\Big)\,, \label{TOmOm}
\end{align}
where 
\begin{align}
(X_{23})_{\nu} = \frac{(x_{12})_{\nu}}{x_{12}^{2}} - \frac{(x_{13})_{\nu}}{x_{13}^{2}} \,.
\end{align}
The conformal Ward identity gives
\begin{align}
C_{T O_{m}O_{m} } = \frac{1}{S_{d}} \frac{d\Delta_{O_{m}}}{d-1}C_{O_{m}}\,, \label{TOOCO}
\end{align}
where $C_{O_{m}}$ and $\Delta_{O_{m}}$ are two-point constant and anomalous dimension of the operator $O_{m}$ in coordinate space:
\begin{align}
\langle O^{\textrm{ren}}_{m}(x)O^{\textrm{ren}}_{m}(0)\rangle  = \frac{C_{O_{m}}}{(x^{2})^{\Delta_{O_{m}}}}\,.
\end{align}
Taking the Fourier transform of (\ref{TOmOm})  and setting the momentum of the stress-energy tensor to zero for simplicity, one finds in terms of the 
projected stress tensor $T=z^{\mu}z^{\nu}T_{\mu\nu}$
\begin{align}
\langle T^{\textrm{ren}}(0) O^{\textrm{ren}}_{m}(p)O^{\textrm{ren}}_{m}(-p) \rangle= (d-2\Delta_{O_{m}})\tilde{C}_{O_{m}} \frac{p_{z}^{2}}{(p^{2})^{\frac{d}{2}-\Delta_{O_{m}}+1}}\,, \label{TOmOmWard}
\end{align}
where $\tilde{C}_{O_{m}}$ is the  two-point constant of $\langle O^{\textrm{ren}}_{m} O^{\textrm{ren}}_{m}\rangle$ correlator in the momentum space: 
 \begin{align}
\langle O^{\textrm{ren}}_{m}(p)O^{\textrm{ren}}_{m}(-p)\rangle  = \frac{\tilde{C}_{O_{m}}}{(p^{2})^{1-\frac{d}{2}-\eta_{m}}}\,,
\end{align}
and  $\Delta_{O_{m}}=d-1+\eta_{m}$, where $\eta_{m}=\eta_{m1}/N+\mathcal{O}(1/N^{2})$. 
In order to find $\tilde{C}_{O_{m}}$, $Z_{O_{m}}$ and $\eta_{m1}$ up to $1/N$ order, we have to calculate the diagrams depicted in figure \ref{OmOmDiag}.
The expressions for the diagrams are
\begin{align}
&D_{0}= \int \frac{d^{d}p_{1}}{(2\pi)^{d}} (-1)\Tr(G(p+p_{1})G(p_{1})) \,, \notag\\
&D_{1}= 2(i)^{2}\mu^{2\Delta} \int \frac{d^{d}p_{1}d^{d}p_{2}}{(2\pi)^{2d}} (-1)\Tr(G(p+p_{1})G(p_{1})\gamma^{\nu_{1}}G(p_{2})\gamma^{\nu_{2}}G(p_{1}))D_{\nu_{1}\nu_{2}}(p_{1}-p_{2})\,,\notag\\
&D_{2}= (i)^{2}\mu^{2\Delta} \int \frac{d^{d}p_{1}d^{d}p_{2}}{(2\pi)^{2d}} (-1)\Tr(G(p+p_{1})\gamma^{\nu_{1}}G(p+p_{2})G(p_{2})\gamma^{\nu_{2}}G(p_{1}))D_{\nu_{1}\nu_{2}}(p_{1}-p_{2})
\end{align}
and 
\begin{align}
\langle O^{\textrm{ren}}_{m}(p)O^{\textrm{ren}}_{m}(-p)\rangle=Z_{O_{m}}^{2}\langle O_{m}(p) O_{m}(-p)\rangle = Z_{O_{m}}^{2}\big(D_{0}+D_{1}+D_{2}+\mathcal{O}(1/N^{2})\big)\,.
\end{align}
Computing these diagrams one finds
\begin{align}
2Z_{O_{m}1}=\eta_{m1}=-\frac{2 (d -1) \Gamma (d )}{\Gamma (\frac{d}{2} )^2 \Gamma (\frac{d}{2} +1) \Gamma (2-\frac{d}{2} )}\, \label{etamap}
\end{align}
and 
\begin{align}
\tilde{C}_{O_{m}}=\frac{4^{1-d } \pi ^{\frac{3-d}{2} } \Tr {\bf 1}}{\Gamma \left(\frac{d-1}{2}\right) \sin (\pi \frac{d}{2} )} \bigg(1+\frac{1}{N}\eta_{m1}\Big(\frac{3  d(d -2)}{8 (d -1)}\Theta (d )-\Psi(d )+\frac{d-2}{ d }\Big)\bigg)\,,
\end{align}
where $\Theta(d)\equiv \psi'(d/2)-\psi'(1)$ and $\Psi(d)\equiv \psi(d-1)+\psi(2-d/2)-\psi(1)-\psi(d/2-1)$. 

Now we can calculate the three-point function  $\langle T^{\textrm{ren}}(0)  O^{\textrm{ren}}_{m}(p) O^{\textrm{ren}}_{m}(-p) \rangle$ using Feynman diagrams, 
namely we have
\begin{align}
\langle T^{\textrm{ren}}(0)  O^{\textrm{ren}}_{m}(p)O^{\textrm{ren}}_{m}(-p) \rangle = Z_{T}Z_{ O_{m}}^{2}\langle T(0)  O_{m}(p) O_{m}(-p) \rangle \label{Tpsipsidiag}
\end{align}
and the diagrams contributing to $\langle T(0)  O_{m}(p) O_{m}(-p)\rangle$ up to order  $1/N$  are shown in figure \ref{TpsipsiQED}, and the explicit 
results are listed in eq.~(\ref{ZTdiag}) below. 
Putting these diagrams together and equating the expression (\ref{TOmOmWard}) required by conformal symmetry with the diagrammatic result for (\ref{Tpsipsidiag}), 
we find that the required $Z_T$ factor is the one given in (\ref{Ztfermap}).
As a check of our calculation, we note that dependence on the gauge parameter $\xi$ drops out from the final result. 

Let us end this section by listing the results for the diagrams in figure \ref{TpsipsiQED}. They are given by
\begin{align}
D_{0}=&-\Tr {\bf 1}\frac{\pi  \csc (\pi  \frac{d}{2} ) \Gamma (  \frac{d}{2}  )}{(4 \pi )^{  \frac{d}{2}  } \Gamma (d -2)} \frac{p_{z}^{2}}{(p^{2})^{2-\frac{d}{2} }}\,, \notag\\
D_{1}=&\frac{1}{N}D_{0}\eta_{m1}\bigg(\Big(\frac{1}{\Delta}-\log(\frac{p^{2}}{\mu^{2}})\Big)\Big(\frac{d -4}{4}+\frac{d  \xi }{4 (d -1)}\Big)+\Big(\Big(\frac{d -4}{4}+\frac{d  \xi }{4 (d -1)}\Big)\Psi (d )\notag\\
&~~~~~~~~~-\frac{d^3-8 d^2+16 d-16}{4 (d-2) d}-\frac{d^2 \xi }{4 (d-2) (d-1)}\Big)\bigg)\,, \notag\\
D_{2}=&\frac{1}{N}D_{0}\eta_{m1}\bigg(-\Big(\frac{1}{\Delta}-\log(\frac{p^{2}}{\mu^{2}})\Big)\Big(\frac{d^3-7 d^2+10 d-8}{8 (d-1) (d+2)}+\frac{d  \xi }{8 (d -1)}\Big)-\Big(\Big(\frac{d^3-7 d^2+10 d-8}{8 (d-1) (d+2)}+ \notag\\
&\frac{d  \xi }{8 (d -1)}\Big)\Psi-\frac{2 d^7-21 d^6+63 d^5-68 d^4-60 d^3+192 d^2-160 d+64}{8 (d-2) (d-1)^2 d (d+2)^2}-\frac{(2 d^3-7 d^2+12 d-8) \xi }{8 (d-2) (d-1)^2}\Big)\bigg)\,, \notag\\
D_{3}=&\frac{1}{N}D_{0}\eta_{m1}\bigg(-\Big(\frac{1}{\Delta}-\log(\frac{p^{2}}{\mu^{2}})\Big)\Big(\frac{d }{4}+\frac{d  \xi }{4 (d -1)}\Big)+\Big(\frac{3 d  (d -2)^2 }{8(d -1)^2}\Theta (d )-\Big(\frac{d }{4}+\frac{d  \xi }{4 (d -1)}\Big)\Psi (d )\notag\\
&~~~~~~~~~~~~+\frac{d^3-d^2+2 d-4}{4 (d-2) (d-1)^2}+\frac{\left(3 d^2-6 d+4\right) \xi }{4 (d-2) (d-1)^2}\Big)\bigg)\,, \notag \\
D_{4}=&\frac{1}{N}D_{0}\eta_{m1}\bigg(\Big(\frac{1}{\Delta}-\log(\frac{p^{2}}{\mu^{2}})\Big)\Big(\frac{d -4}{8}+\frac{d  \xi }{8 (d -1)}\Big)+\Big(\Big(\frac{d -4}{8}+\frac{d  \xi }{8 (d -1)}\Big) \Psi (d )\notag\\
&~~~~~~~~~~~+\frac{3 d^3-16 d^2+32 d-16}{8 (d-2) (d-1) d}-\frac{d^2 \xi }{8 (d-2) (d-1)^2}\Big)\bigg)\,, \notag\\
D_{5}=&\frac{1}{N}D_{0}\eta_{m1}\bigg(-\Big(\frac{1}{\Delta}-2\log(\frac{p^{2}}{\mu^{2}})\Big)\Big(\frac{(d-2)^2}{4 (d-1) (d+2)}\Big)-\Big(\frac{(d-2)^2}{2 (d-1) (d+2)}\Psi (d )\notag\\
&~~~~~~~~~~~-\frac{(d-2) \left(5 d^4-9 d^3+4 d^2+28 d-16\right)}{4 (d-1)^2 d (d+2)^2}-\frac{(d -2) \xi }{2(d -1)^2}\Big)\bigg)\,, \notag \\
D_{6}=&\frac{1}{N}D_{0}\eta_{m1}\bigg(\frac{3 d(d -2)  }{8 (d -1)^2} \Theta (d )+\frac{d -2}{4 (d -1)}-\frac{(d -2) \xi }{2(d -1)^2}\bigg)\,, \notag\\
D_{7}=&\frac{1}{N}D_{0}\eta_{m1}\bigg(\frac{3 d(d -2)   }{8 (d -1)^2}\Theta (d )+\frac{1}{2 (d -1)}-\frac{\xi }{2 (d -1)}\bigg)\,, \notag\\
D_{8}=&\frac{1}{N}D_{0}\eta_{m1}\bigg(\Big(\frac{1}{\Delta}-\log(\frac{p^{2}}{\mu^{2}})\Big)\Big(\frac{d -2}{2(d -1)}\Big)+\Big(\frac{(d -2) }{2(d -1)}\Psi (d )-\frac{d^2-3 d+4}{2 (d-1) d}+\frac{\xi }{2 (d -1)}\Big)\bigg)\,, \notag\\
D_{9}=&\frac{1}{N}D_{0}\eta_{m1}\bigg(-\Big(\frac{1}{\Delta}-2\log(\frac{p^{2}}{\mu^{2}})\Big)\Big(\frac{d -2}{4 (d -1)}\Big)-\Big(\frac{(d -2) }{2(d -1)}\Psi (d )-\frac{d^3-3 d^2+5 d-4}{2 (d-1)^2 d}+\frac{(d -2) \xi }{2(d -1)^2}\Big)\bigg)\,, \notag\\
D_{10}=&\frac{1}{N}D_{0}\eta_{m1}\bigg(-\frac{3 d(d -2)   }{8 (d -1)^2}\Theta (d )-\frac{2d -3}{2 (d -1)^2}+\frac{(d -2) \xi }{2(d -1)^2}\bigg)\,,
\label{ZTdiag}
\end{align}
where $\Theta(d)\equiv \psi'(d/2)-\psi'(1)$ and $\Psi(d)\equiv \psi(d-1)+\psi(2-d/2)-\psi(1)-\psi(d/2-1)$ and $\eta_{m1}$ is given in (\ref{etamap}).
We notice that 
\begin{align}
D_{7}+D_{8}+D_{9}+D_{10}=\frac{D_{0}\eta_{m1}}{N\Delta}\frac{ (d -2)}{4(d -1)}\,.
\end{align}

\begin{figure}[h!]
                \centering
                \includegraphics[width=17cm]{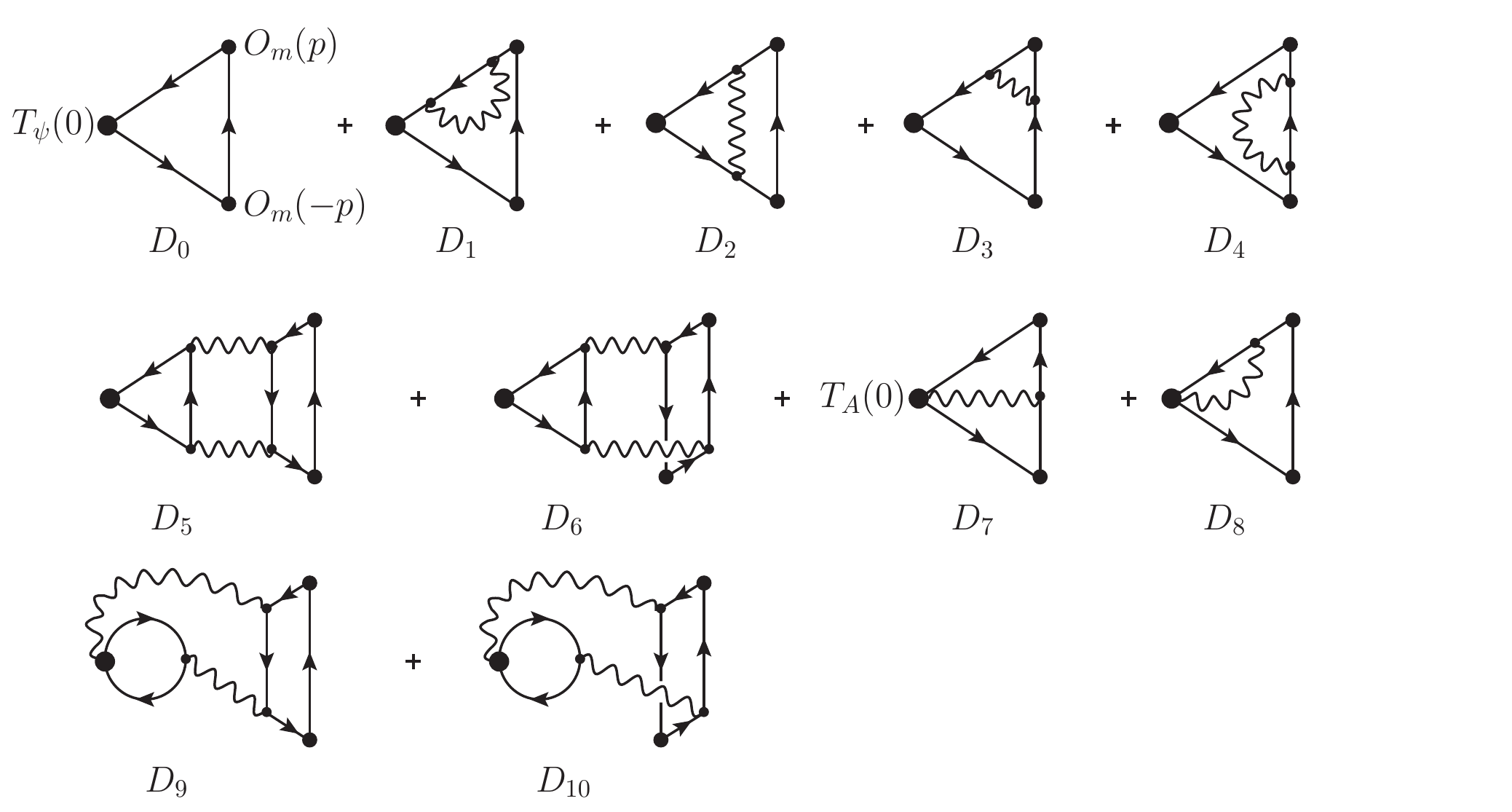}
                \caption{Diagrams contributing to $\langle T(0)  O_{m}(p) O_{m}(-p) \rangle$ up to order $1/N$. }
                \label{TpsipsiQED}
\end{figure}

\section{Results for $\langle J J\rangle$  and $\langle T T\rangle$ diagrams} \label{apb}
The diagrams for $\langle JJ  \rangle $ shown in figure \ref{CJQED} are given explicitly by 
\begin{align}
&D_{0}= \tr (t^{a}t^{b}) \int \frac{d^{d}p_{1}}{(2\pi)^{d}} (-1)\Tr (\gamma_{z}G(p_{1})\gamma_{z}G(p+p_{1}))\,, \notag\\
&D_{1}= 2\tr (t^{a}t^{b})(i)^{2} \mu^{2\Delta}\int \frac{d^{d}p_{1}d^{d}p_{2}}{(2\pi)^{2d}} (-1)\Tr (\gamma_{z}G(p+p_{1})\gamma_{z}G(p_{1})\gamma^{\nu_{1}}G(p_{2})\gamma^{\nu_{2}}G(p_{1}))D_{\nu_{1}\nu_{2}}(p_{1}-p_{2}) \,,\notag\\
&D_{2}= \tr (t^{a}t^{b})(i)^{2}\mu^{2\Delta} \int \frac{d^{d}p_{1}d^{d}p_{2}}{(2\pi)^{2d}} (-1)\Tr (\gamma_{z}G(p+p_{1})\gamma^{\nu_{1}}G(p+p_{2})\gamma_{z}G(p_{2})\gamma^{\nu_{2}}G(p_{1}))D_{\nu_{1}\nu_{2}}(p_{1}-p_{2})
\end{align}
and the results are 
\begin{align}
&D_{0}= \tr (t^{a}t^{b})\Tr {\bf 1}\frac{ \pi \csc \left(\frac{\pi  d}{2}\right) \Gamma \left(\frac{d}{2}\right)}{(4\pi)^{\frac{d}{2}}(d-1) \Gamma (d-2)} \frac{p_{z}^{2}}{(p^{2})^{2-\frac{d}{2}}} \,, \notag\\
&D_{1}=\frac{1}{N}D_{0}\eta_{m1} \bigg( \Big(\frac{1}{\Delta }-\log (\frac{p^2}{\mu^{2}})\Big)\Big(\frac{d-4}{4}+\frac{d \xi }{4 (d-1)}\Big)+\Big(\Big(\frac{d-4}{4}+\frac{d \xi }{4 (d-1)}\Big) \Psi(d) \notag\\
&~~~~~~~~+\frac{3 d^3-16 d^2+32 d-16}{4 (d-2) (d-1) d}-\frac{d^2 \xi }{4 (d-2) (d-1)^2}\bigg) \bigg ) \,,\notag\\
&D_{2}= \frac{1}{N}D_{0}\eta_{m1} \bigg(- \Big(\frac{1}{\Delta }-\log (\frac{p^2}{\mu^{2}})\Big)\Big(\frac{d-4}{4}+\frac{d \xi }{4 (d-1)}\Big)+\Big(-\Big(\frac{d-4}{4}+\frac{d \xi }{4 (d-1)}\Big) \Psi(d) \notag\\
&~~~~~~~~+\frac{3 d(d-2) }{8 (d-1)} \Theta (d)+\frac{(d-4) d}{4 (d-2) (d-1)}+\frac{d^2 \xi }{4 (d-2) (d-1)^2}\bigg)  \bigg )\,.
\end{align}

The diagrams for  $\langle T T\rangle$ depicted in figure \ref{CTQEDdiag} are given explicitly by 
\begin{align}
D_{0} =&N_{f} \big(\frac{-i}{2}\big)^{2} \int \frac{d^{d}p_{1}}{(2\pi)^{d}} (2p_{1z}+p_{z})^{2}(-1)\Tr (\gamma_{z}G(p+p_{1})\gamma_{z}G(p_{1})),\notag\\
D_{1}=&2N_{f}\mu^{2\Delta}\big(\frac{-i}{2}\big)^{2}(i)^{2}\int \frac{d^{d}p_{1}d^{d}p_{2}}{(2\pi)^{2d}}(2p_{1z}+p_{z})^{2}(-1) \Tr(\gamma_{z} G(p+p_{1})\gamma_{z}G(p_{1})\gamma^{\nu_{1}}G(p_{2})\gamma^{\nu_{2}}G_{p_{1}})D_{\nu_{1}\nu_{2}}(p_{1}-p_{2})\,, \notag \\
D_{2}=&N_{f}\mu^{2\Delta}\big(\frac{-i}{2}\big)^{2}(i)^{2}\int \frac{d^{d}p_{1}d^{d}p_{2}}{(2\pi)^{2d}}(2p_{1z}+p_{z})(2p_{2z}+p_{z})(-1) \Tr(\gamma_{z} G(p+p_{1})\gamma^{\nu_{1}}G(p+p_{2})\gamma_{z}G(p_{2})\gamma^{\nu_{2}}G_{p_{1}})\notag\\
&\times D_{\nu_{1}\nu_{2}}(p_{1}-p_{2})\,, \notag \\
D_{3}=&N_{f}^{2}\mu^{4\Delta}\big(\frac{-i}{2}\big)^{2}(i)^{4}\int \frac{d^{d}p_{1}d^{d}p_{2}d^{d}p_{3}}{(2\pi)^{3d}}(2p_{1z}+p_{z})(-1) \Tr(\gamma_{z} G(p+p_{1})\gamma^{\nu_{1}}G(p_{1}-p_{3})\gamma^{\nu_{2}}G(p_{1}))\notag\\
&\times D_{\nu_{1}\nu_{3}}(p+p_{3})D_{\nu_{2}\nu_{4}}(p_{3}) (2p_{2z}+p_{z})(-1) \Tr(\gamma_{z} G(p_{2})\gamma^{\nu_{4}}G(p_{2}-p_{3})\gamma^{\nu_{3}}G(p+p_{2}))+\dots\,, \notag\\
D_{4}=&2N_{f}\mu^{2\Delta}\big(\frac{-i}{2}\big)(-i)(i)\int \frac{d^{d}p_{1}d^{d}p_{2}}{(2\pi)^{2d}}(2p_{1z}+p_{z})(-1)\Tr (\gamma_{z}G(p+p_{1})\gamma_{z}G(p_{2})\gamma^{\nu_{1}}G(p_{1}))D_{\nu_{1}z}(p_{1}-p_{2})\,, \notag\\
D_{5}=&2N_{f}^{2}\mu^{4\Delta}\big(\frac{-i}{2}\big)(-i)(i)^{3}\int \frac{d^{d}p_{1}d^{d}p_{2}d^{d}p_{3}}{(2\pi)^{3d}}(-1)\Tr (\gamma_{z}G(p_{1}-p_{3})\gamma^{\nu_{1}}G(p_{1}))D_{\nu_{1}\nu_{2}}(p_{3})D_{z\nu_{3}}(p+p_{3}) \notag\\
&\times (2p_{2z}+p_{z})(-1) \Tr (\gamma_{z}G(p_{2})\gamma^{\nu_{2}}G(p_{2}-p_{3})\gamma^{\nu_{3}}G(p+p_{2}))+\dots\,, \notag\\
D_{6}=&N_{f}\mu^{2\Delta}(-i)^{2}\int \frac{d^{d}p_{1}d^{d}p_{2}}{(2\pi)^{2d}}(-1)\Tr (\gamma_{z}G(p+p_{2})\gamma_{z}G(p_{1}))D_{zz}(p_{1}-p_{2})\,, \notag\\
D_{7}=&N_{f}^{2}\mu^{4\Delta}(-i)^{2}(i)^{2}\int \frac{d^{d}p_{1}d^{d}p_{2}d^{d}p_{3}}{(2\pi)^{3d}}(-1)\Tr (\gamma_{z}G(p_{1}-p_{3})\gamma^{\nu_{1}}G(p_{1}))D_{zz}(p+p_{3})D_{\nu_{1}\nu_{2}}(p_{3})\notag\\
&\times (-1)\Tr(\gamma_{z}G(p_{2})\gamma^{\nu_{2}}G(p_{2}-p_{3}))\,, \notag\\
D_{8}=&N_{f}^{2}\mu^{4\Delta}(-i)^{2}(i)^{2}\int \frac{d^{d}p_{1}d^{d}p_{2}d^{d}p_{3}}{(2\pi)^{3d}}(-1)\Tr (\gamma_{z}G(p_{1}-p_{3})\gamma^{\nu_{1}}G(p_{1}))D_{\nu_{1}z}(p_{3})D_{z\nu_{2}}(p+p_{3})\notag\\
&\times (-1)\Tr(\gamma_{z}G(p_{2}-p_{3})\gamma^{\nu_{2}}G(p+p_{2}))\,
\end{align}
where dots mean that there is also an expression which corresponds to the opposite direction of the fermion loop. After carrying out the momentum integrals 
using techniques similar to the ones described in the appendices of \cite{Diab:2016spb}, we find
\begin{align}
D_{0}=&-N\frac{\pi ^{1-\frac{d}{2} } \csc (\pi \frac{d}{2} ) \Gamma (\frac{d}{2} )}{4^{\frac{d}{2} +1} (d-1) (d +1) \Gamma (d -2)}\frac{p_{z}^{4}}{(p^{2})^{2-\frac{d}{2}}}\,,\notag\\
D_{1}= &\frac{1}{N}D_{0}\eta_{m1}\bigg(\Big(\frac{1}{\Delta}-\log(\frac{p^{2}}{\mu^{2}})\Big)\Big(\frac{d -4}{4}+\frac{d  \xi }{4 (d -1)}\Big)+\Big(\Big(\frac{d -4}{4}+\frac{d  \xi }{4 (d -1)}\Big)\Psi(d )\notag\\
&~~~~~~~~~~+\frac{2 d^4-10 d^3+15 d^2+4 d-8}{2 (d-2) (d-1) d (d+1)}-\frac{d (2 d-1) \xi }{2 (d-2) (d-1)^2 (d+1)}\Big)\bigg)\,, \notag\\
D_{2}= &\frac{1}{N}D_{0}\eta_{m1}\bigg(-\Big(\frac{1}{\Delta}-\log(\frac{p^{2}}{\mu^{2}})\Big)\Big(\frac{d^3-7 d^2+10 d-8}{4 (d-1) (d+2)}+\frac{d  \xi }{4 (d -1)}\Big)+\Big(\frac{3 d(d -2)  }{8 (d -1)} \Theta(d )\notag\\
&~~~~~~~~~~-\Big(\frac{d^3-7 d^2+10 d-8}{4 (d-1) (d+2)}+\frac{d  \xi }{4 (d -1)}\Big)\Psi(d )+\frac{5 d^5-27 d^4+44 d^3-30 d^2-12 d+16}{2 (d-2) (d-1)^2 d (d+1) (d+2)}\notag\\
&~~~~~~~~~~-\frac{\left(d^3-4 d^2+2\right) \xi }{2 (d-2) (d-1)^2 (d+1)}\Big)\bigg)\,, \notag\\
D_{3}= &\frac{1}{N}D_{0}\eta_{m1}\bigg(-\Big(\frac{1}{\Delta}-2\log(\frac{p^{2}}{\mu^{2}})\Big)\Big(\frac{(d-2)^2}{2 (d-1) (d+2)}\Big)+\Big(\frac{2 (d-2)}{(d-1) (d+2)}\Psi(d )\notag\\
&~~~~~~~~~~-\frac{d \left(d^3-8 d+11\right)}{(d-1)^2 (d+1) (d+2)}+\frac{\xi }{2 (d -1)}\Big)\bigg)+d D_{0}/(2N)\,, 
\end{align}
and 
\begin{align}
D_{4}= &\frac{1}{N}D_{0}\eta_{m1} \bigg(\Big(\frac{1}{\Delta}-\log(\frac{p^{2}}{\mu^{2}})\Big)\Big(\frac{d -2}{d -1}\Big)+\Big(\frac{ (d -2)\Psi(d) }{d -1}-\frac{d^4-4 d^3+5 d^2+2 d-2}{(d-1)^2 d (d+1)}+\frac{\xi }{d-1 }\Big)\bigg)\,, \notag\\
D_{5}= &\frac{1}{N}D_{0}\eta_{m1} \bigg( -\Big(\frac{1}{\Delta}-2\log(\frac{p^{2}}{\mu^{2}})\Big)\Big(\frac{d -2}{2(d -1)}\Big)-\Big(\frac{ (d -2)\Psi(d)}{d -1}-\frac{d^4-4 d^3+5 d^2+2 d-2}{(d-1)^2 d (d+1)}+\frac{\xi }{d -1}\Big)\bigg)\notag\\
&-d D_{0}/N\,, \notag\\
D_{6}= &\frac{1}{N}D_{0}\eta_{m1} \bigg( -\frac{\xi -1}{2 (d -1)}\bigg)\,, \notag\\
D_{7}= &\frac{1}{N}D_{0}\eta_{m1} \bigg(\frac{\xi -1}{2 (d -1)} \bigg)\,, \notag\\
D_{8}= &d D_{0}/(2N)\,,
\end{align}
where $\Theta(d)\equiv \psi'(d/2)-\psi'(1)$ and $\Psi(d)\equiv \psi(d-1)+\psi(2-d/2)-\psi(1)-\psi(d/2-1)$ and $\eta_{m1}$ is given in (\ref{etamap}).
We notice that 
\begin{align}
D_{4}+D_{5}+D_{6}+D_{7}+D_{8}= \frac{D_{0}\eta_{m1}}{N\Delta}\frac{(d-2)}{2(d-1)}- d D_{0}/(2N) \,.
\end{align}

\bibliographystyle{ssg}
\bibliography{ctqed}

\end{document}